\documentclass{iaus}
\usepackage{graphicx}

\newcommand{\lsim}{\ \raise -2.truept\hbox{\rlap{\hbox{$\sim$}}\raise5.truept
        \hbox{$<$}\ }}
\newcommand{\gsim}{\ \raise -2.truept\hbox{\rlap{\hbox{$\sim$}}\raise5.truept
        \hbox{$>$}\ }} 
\newcommand{\ngrbs}{150}
\newcommand{\nhosts}{46}
\newcommand{\ndlas}{37}

\title[JD 11.~~GRB Host Galaxies] 
{Low-Mass and Metal-Poor Gamma-Ray Burst Host Galaxies}

\author[Sandra Savaglio]   
{Sandra Savaglio$^1$}

\affiliation{$^1$Max-Planck Institute for Extraterrestrial Physics, \\ Giessenbachstr.,
PF 1312, D-85741, Garching bei M\"unchen, Germany\\ email: {\tt savaglio@mpe.mpg.de}}

\pubyear{2008}
\volume{255}  
\pagerange{119--126}
\jname{Low-Metallicity Star Formation: From the First Stars to Dwarf Galaxies}
\editors{L.K. Hunt, S. Madden \& R. Schneider, eds.}
\begin{document}

\maketitle

\begin{abstract}
Gamma-ray bursts (GRBs) are cosmologically distributed, very energetic and very transient sources detected in the $\gamma$-ray domain. The identification of their $x$-ray and optical afterglows allowed so far the redshift measurement of  \ngrbs\ events, from $z=0.01$ to $z=6.29$. For about half of them, we have some knowledge of the properties of the parent galaxy.  At high redshift ($z>2$), absorption lines in the afterglow spectra give information on the cold interstellar medium in the host. At low redshift ($z<1.0$) multi-band optical-NIR photometry and integrated spectroscopy reveal the GRB host general properties. A redshift evolution of metallicity is not noticeable in the whole sample. The typical value is a few times lower than solar. The mean host stellar mass is similar to that of the Large Magellanic Cloud, but the mean star formation rate is five times higher. GRBs are discovered with $\gamma$-ray, not optical or NIR, instruments. Their hosts do not suffer from the same selection biases of typical galaxy surveys. Therefore, they might represent a fair sample of the most common galaxies that existed in the past history of the universe, and can be used to better understand galaxy formation and evolution.

\keywords{Galaxies: ISM, galaxies: abundances, galaxies: dwarf, galaxies:  high redshift, gamma rays: bursts, gamma rays: observations}
\end{abstract}

\firstsection 
\section{Introduction}

\begin{figure}
\centerline{\includegraphics[scale = .4]{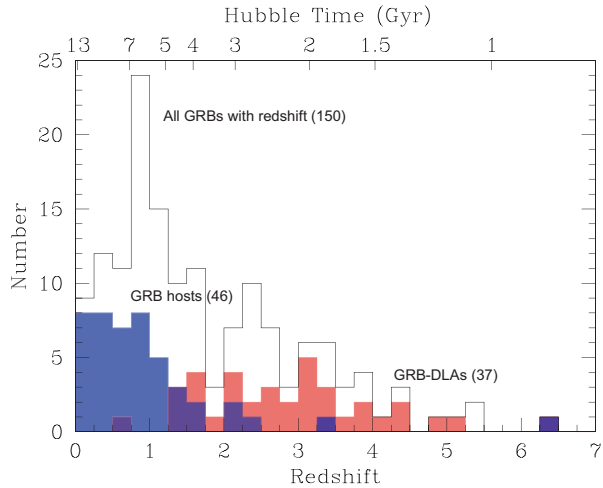}}
\begin{center}
 \caption{Histogram of GRBs with measured redshift (empty histogram \ngrbs\ objects). The high-$z$ and low-$z$ filled histograms are the subsample of GRBs studied using optical afterglow spectra (GRB-DLAs) and multiband photometry of the host galaxies (GRB hosts), respectively.}
   \label{z_hist}
\end{center}
\end{figure}

More than 40 years have passed since the discovery of the first gamma-ray burst (GRB), the most energetic explosions in the universe, by the US military satellite Vela. It took 30 years to demonstrate their cosmological origin (\cite[Metzger et al.\ 1997]{metzger97}). Their $\gamma$-ray energies (typically $10^{51}$ ergs, emitted in less than a couple of minutes) emerge from a collimated jet in a core-collapse supernovae, or the merger of two compact objects (neutron stars or black holes).

Due to the highly transient nature of GRBs, their redshift, measured from the optical afterglow or the host galaxy, is known today for \ngrbs\ objects only (Figure~\ref{z_hist}). Although GRBs are very rare (a rate of 1 event every $10^5$ years in a galaxy is estimated, after correcting for the jet opening angle), they are so energetic that a few events are expected to be detectable from Earth every day. As shown by the discovery of high redshift events (the highest ever being GRB~050904 at $z=6.3$; Kawai et al.\ 2006), GRBs offer the opportunity to explore the most remote universe, under extreme conditions, hard to observe using traditional tools.

The first scientific paper on GRBs dates from 6 years after the Vela discovery (Klebesadel, Strong \& Olson 1973).
In 1975, already 100 different theories where proposed, and today more than 5000 refereed papers on GRBs have been published.  Before the 1997 discovery, theories on the energetic emission ranked from the impact of comets onto neutron stars (Harwit \& Salpeter 1973) to collisions of chunks of antimatter with normal stars (Sofia \& van Horn 1974). However, the most likely hypothesis was already postulated years earlier, when nothing was really known about GRBs: Colgate (1968) predicted $\gamma$-ray emission from supernovae in distant galaxies.

The curse and blessing of GRBs is their fast fading. The light curve is extremely steep and therefore very hard to catch. The emitted energy is so immense that it cannot last long. However, it allows us to see a hidden universe. One extreme case is the recent event GRB~080319B at $z=0.937$ (Bloom et al.\ 2008), visible for a short time by naked eye (pick optical magnitude $m=5.6$). This GRB was already hard to observe spectroscopically with the largest telescopes 7 hours after its discovery.

\begin{table}
  \begin{center}
  \caption{Overview of host diagnostics with GRB observations.}
  \label{t1}
 {\scriptsize
  \begin{tabular}{|l|c|c|c|c|}\hline 
& {\bf GRB-DLA} & {\bf GRB host} & {\bf GRB host} \\ 
{\bf Diagnosis}    &  {\bf (AG spectroscopy)}  & {\bf (Int. spectroscopy)} & {\bf (Photometry)} \\ \hline
SFR & $\times$ & $\surd\surd$ & $\surd$  \\  \hline
Metallicity & $\surd\surd\surd$ & $\surd\surd$ & $\times$  \\ \hline
Dust extinction & $\surd$ & $\surd\surd$& $\times$ \\ \hline
Dust depletion & $\surd\surd\surd$ & $\times$ & $\times$ \\ \hline
 Stellar mass & $\times$ & $\times$ & $\surd\surd\surd$   \\ \hline
Age & $\times$ & $\surd$ & $\surd$ \\ \hline
  \end{tabular}
  }
 \end{center}
\end{table}

\section{Studying small star-forming galaxies at low and high redshift with GRBs}

GRB studies offer the opportunity to complement our partial view of the universe, and better understand the formation and evolution of galaxies in general. The typical GRB host is a small star-forming galaxy (Le Floc'h et al.\ 2003; Christensen et al.\ 2004; Prochaska et al.\ 2004). Galaxies with similar characteristics, but without known GRB,  are dominating the universe. This effect can only have been more important in the past, because galaxies become bigger with time and not smaller. 

High redshift galaxies have been explored for more than two decades using absorption lines in QSO spectra. The so-called damped Lyman-$\alpha$ systems (DLAs) probe the physical state of neutral gas in galaxies intersecting QSO sight lines.  Although very efficient (QSOs are bright and detected up to $z=6.4$), this technique does not allow, except for some special cases, the study of the emitting component (stars and hot gas) of the parent galaxies because of the strong glare of the background light. 

The direct detection of a large number of high-$z$ galaxies was possible with the advent of modern technologies of 10m-class and space telescopes. The Lyman-break technique (Steidel et al.\ 1996) revealed that massive and chemically evolved galaxies where already in place up to  $z=3$ (Erb et al.\ 2006). The difference with QSO-DLA galaxies (metal poor and small) is the effect of Lyman-break galaxies being the tip of the iceberg of the whole galaxy population. The typical high-$z$ galaxy is much fainter and metal poor, and also much harder to find. Much harder, unless a GRB event takes place.

The recent exploration of the SFR density of the universe for different stellar mass bins (Juneau et al.\ 2005) has shown that small galaxies (like GRB hosts) prevailed in terms of star-formation in the $z<2$ universe. Moreover, the mass-metallicity relation (Tremonti et al.\ 2004) and its redshift evolution (Savaglio et al. 2005; Erb et al.\ 2006; Maiolino et al.\ 2008) have revealed that small galaxies have evolved more and over a longer time than big galaxies. Therefore, it is important to go deeper in terms of stellar mass to understand galaxy formation in the distant universe. GRB hosts can serve this purpose. Last but not least, using GRB observations it is possible to study the cold gas and the emitting stars and gas in the same galaxy.

Table~\ref{t1} summarizes the physical parameters studied with GRBs. Observables are grouped in three categories: $i)$ the spectroscopy of the optical afterglow (AG); $ii)$ the integrated optical spectroscopy of the host galaxy; $iii)$ the multi-band optical and NIR photometry of the host. The diagnosis that can (marked with $\surd$ ) or cannot (marked with $\times$)  be investigated are listed. In this contribution we will describe most of them (but the age of the stellar population and the dust extinction in the host).

\begin{figure}
\centerline{\includegraphics[scale = .6]{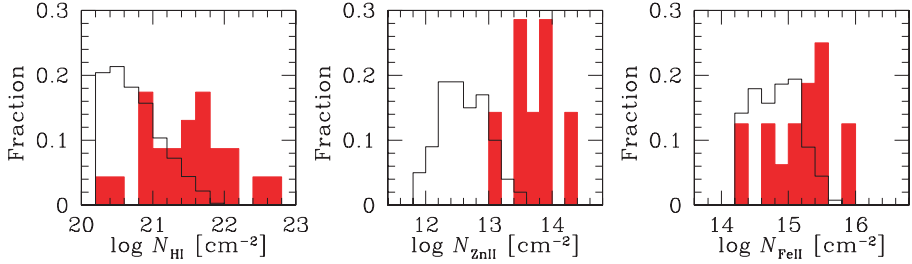}}
\begin{center}
 \caption{Fraction of QSO-DLAs (empty histograms) or GRB-DLAs (filled histograms) per column density bins for different ions (from left to right: HI, ZnII and FeII).}
   \label{fig2}
\end{center}
\end{figure}

\section{Spectroscopy of optical afterglows}

Starting from discovery of the first optical GRB afterglow (Metzger et al.\ 1997), it became clear that GRB spectra looked similar to QSO-DLAs (Savaglio, Fall \& Fiore 2003), with the presence of strong absorption lines of neutral or singly ionized species. The main difference is that GRB strong absorption lines (GRB-DLAs) probe the gas in the GRB vicinity and inside the host galaxy, while most QSO-DLAs are physically unrelated to the background source. 

GRB-DLAs have on average higher column densities of absorption lines than QSO-DLAs. This is shown in
Figure~\ref{fig2}, with the histograms of the neutral hydrogen and the singly ionized zinc and iron. The difference is even larger if one considers that the column density of a species is higher when the background source is external to the galaxy (as for QSO-DLAs) than when the background source is inside the galaxy  (as for GRB-DLAs). In the former case, the line of sight crosses the entire galaxy, whereas in the latter only the gas in front of the stars is probed. The simplest situation would be that the GRB is at the center and therefore the observed column is a factor of two less than what a background 
source would see.

\begin{figure}
\centerline{\includegraphics[scale = .6]{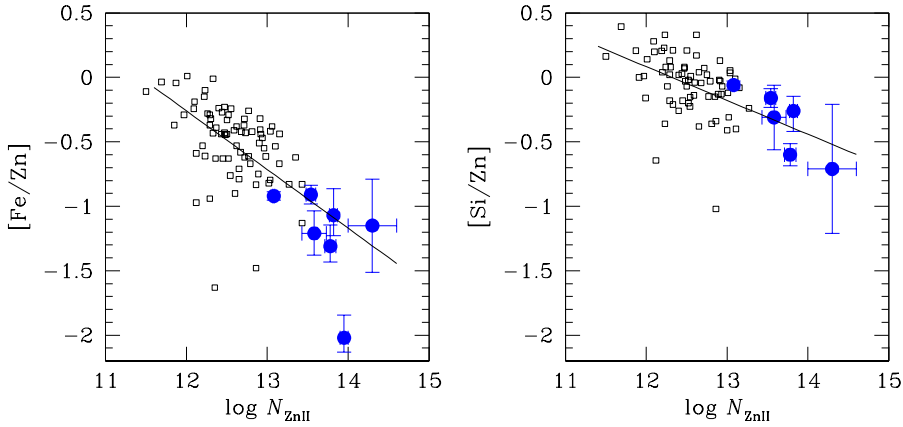}}
\begin{center}
 \caption{Iron-to-zinc and silicon-to-zinc relative abundances as compared to ZnII column density, for QSO-DLAs (open squares) and GRB-DLAs (filled dots). The straight lines show the linear correlations between quantities.}
   \label{fig3}
\end{center}
\end{figure}

As of today, one or more column densities of elements is known for \ndlas\ GRB-DLAs (Figure~\ref{z_hist}; Savaglio 2006 and references therein). The larger column densities with respect to QSO-DLAs can be explained if GRB hosts are larger in size and/or the gas along the sight lines is denser than in QSO-DLAs. Moreover, from a close look at the column density distributions (Figure~\ref{fig2}), it can be noticed that the difference between QSO-DLAs and GRB-DLAs is stronger for ZnII than for FeII. This can be an indication of a larger dust content in GRB-DLAs. In fact, zinc is little dust depleted, while a large fraction of iron is generally locked into dust grains (Savage \& Sembach 1996); the fraction of iron depleted is larger for larger gas column densities.

This is apparent when comparing [Fe/Zn] and [Si/Zn] to the ZnII column densities (Figure~\ref{fig3}). Silicon is mildly depleted in dust grains, more than iron and less than zinc. Negative and large values of [Fe/Zn] and [Si/Zn], together with large column densities of ZnII, indicate large columns of dust in GRB-DLA sight lines (Savaglio 2006). The continuous distribution of points in QSO-DLAs and GRB-DLAs in Figure~\ref{fig3} suggests that the larger column density of metals in GRB-DLAs is mostly the effect of something intrinsic to the gas (e.g.\ larger gas density) and not to the galaxies (e.g.\ larger gas clouds). 

\begin{figure}
\centerline{\includegraphics[scale = .35]{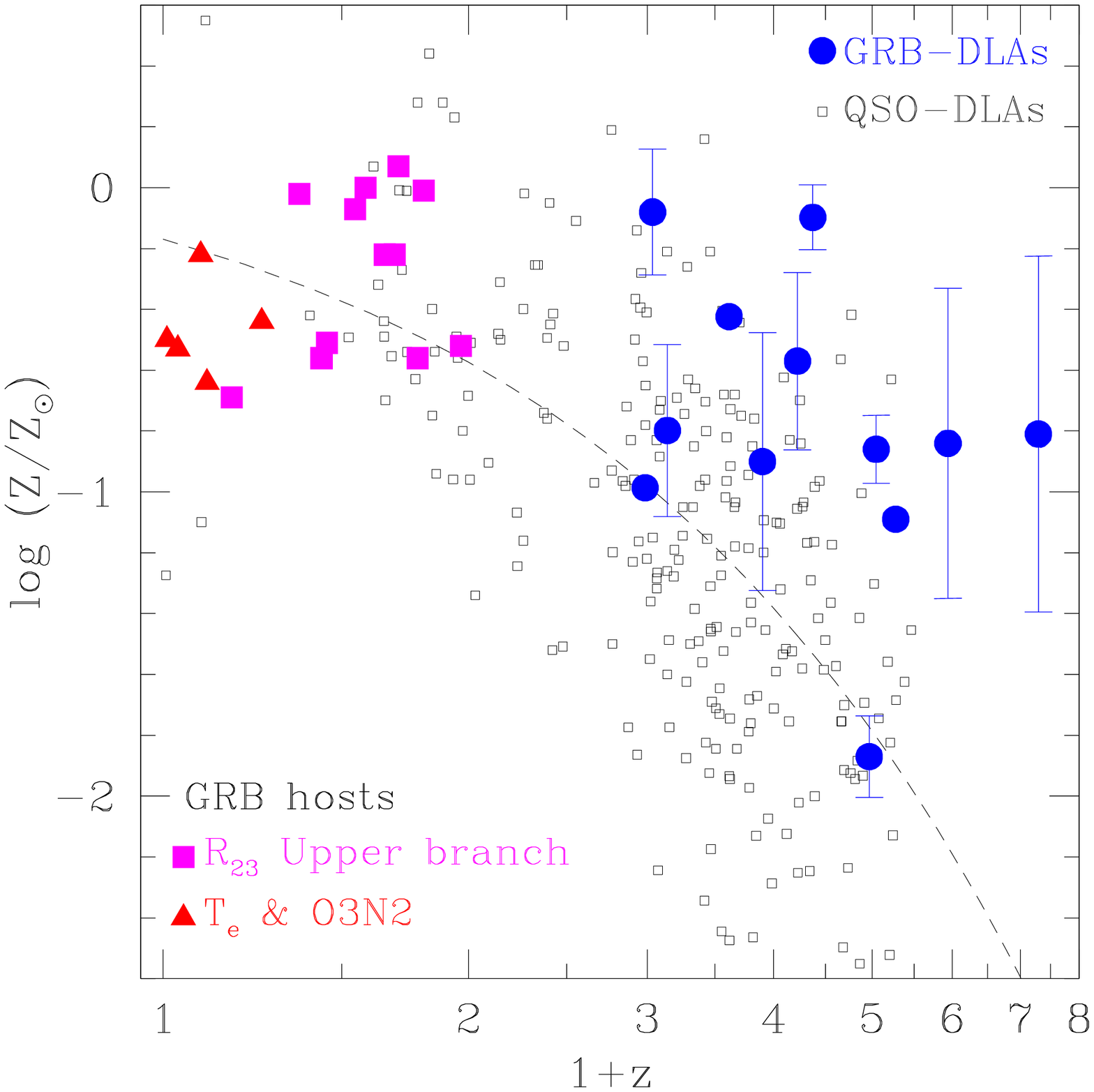}\includegraphics[scale = .35]{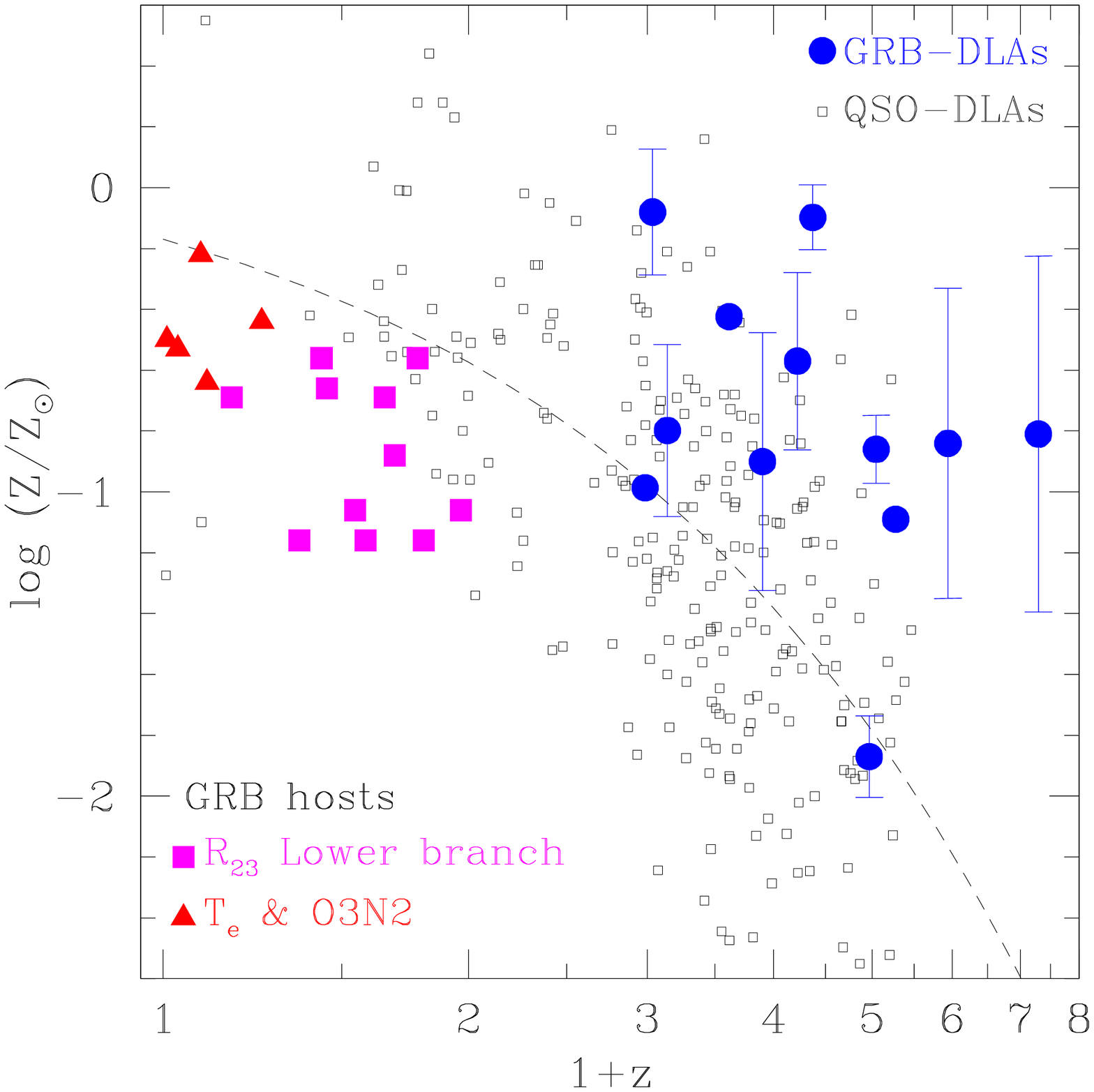}}
\begin{center}
 \caption{Metallicity as a function of redshift for 12 GRB-DLAs (filled circles), 17 GRB hosts (filled triangles and squares), and QSO-DLAs (small open squares). Left and right panels show the upper or lower branch solutions for a subsample of GRB hosts with $R_{23}$ measurements, respectively (Savaglio et al.\ 2008). The dashed line is the fit for QSO-DLAs.}
   \label{fig4}
\end{center}
\end{figure}

The detection of several heavy elements and HI allows to estimate the metal content in GRB-DLAs. Two main assumptions are generally made. First, the ionization correction is neglected (the gas is mostly in neutral form). Second, the dust depletion correction is estimated when more than one element, with different refractory properties, are detected. The metal content in GRB-DLAs at different redshift can be used to trace the chemical evolution of the universe. This was done in the recent years by several authors (Berger et al.\ 2006; Fynbo et al.\ 2006; Savaglio 2006; Prochaska et al.\ 2007). 

The most recent analysis of the metallicity as a function of redshift for 12 GRB-DLAs is shown in Figure~\ref{fig4}. On average GRB-DLAs have larger metallicities than QSO-DLAs. No redshift evolution is detected in the redshift interval $2<z<6.3$. According to the approach followed by Prochaska et al.\ (2007), where several GRB-DLA metallicities are turned into lower limits, the difference is even larger.
 
The source of this significant difference in chemical enrichment is not well understood. Dust obscuration can play a role, as galaxies with dust content as large as those measured in GRB sight lines (Kann et al.\ 2006) would highly obscure background QSOs (Savaglio 2006). Another possibility is the different impact parameter in the two classes of absorbers. Using numerical simulations of star-forming galaxies (Sommer-Larsen \& Fynbo 2008), Fynbo et al.\ (2008) proposed that the difference can be explained by a metallicity gradient in the absorbing galaxies. GRBs are associated with star-forming regions, which are likely located closer to the galaxy center (where metallicity is higher) than the randomly distributed QSO-DLAs.

\begin{figure}
\centerline{\includegraphics[scale = .38]{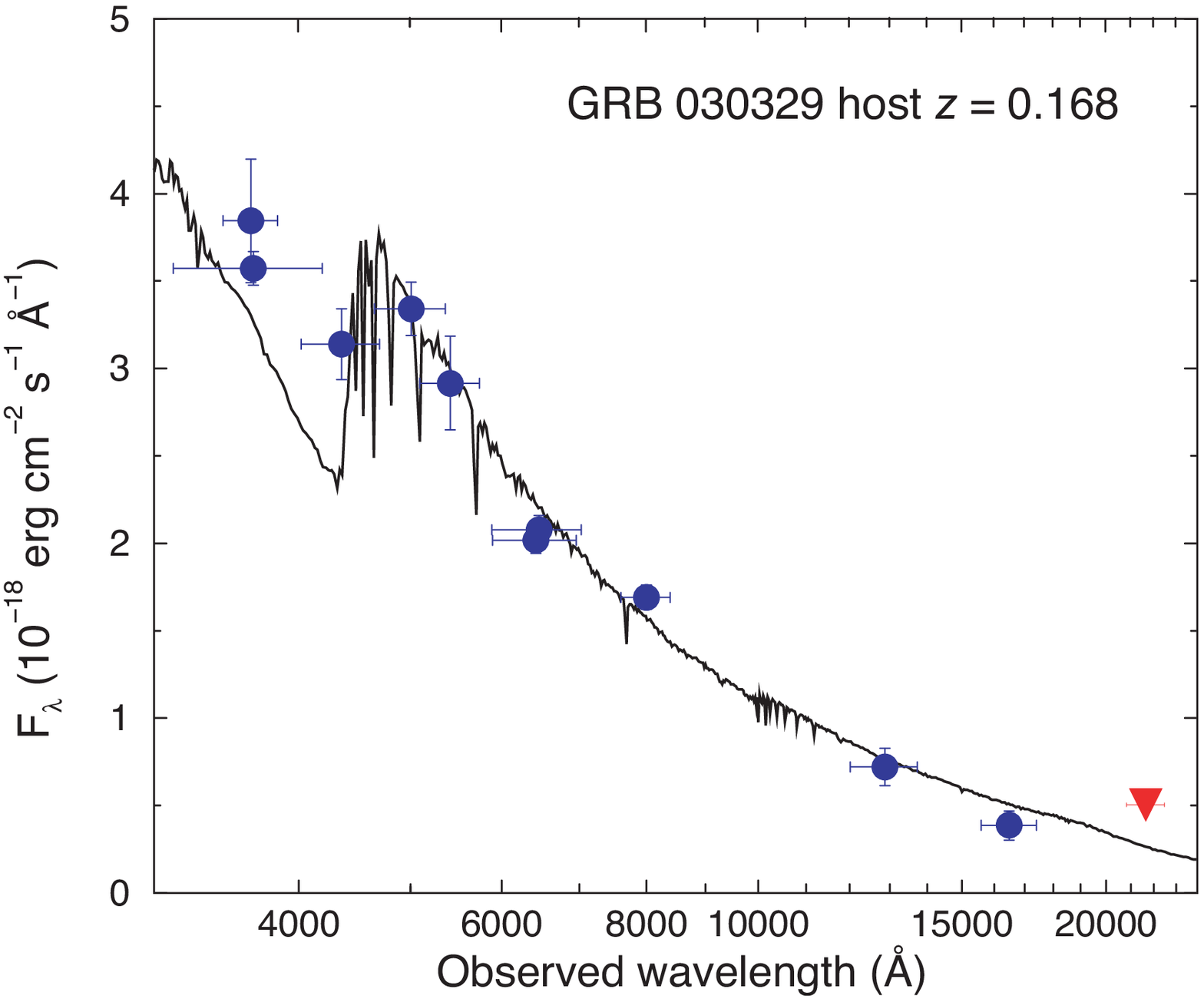}\includegraphics[scale = .62]{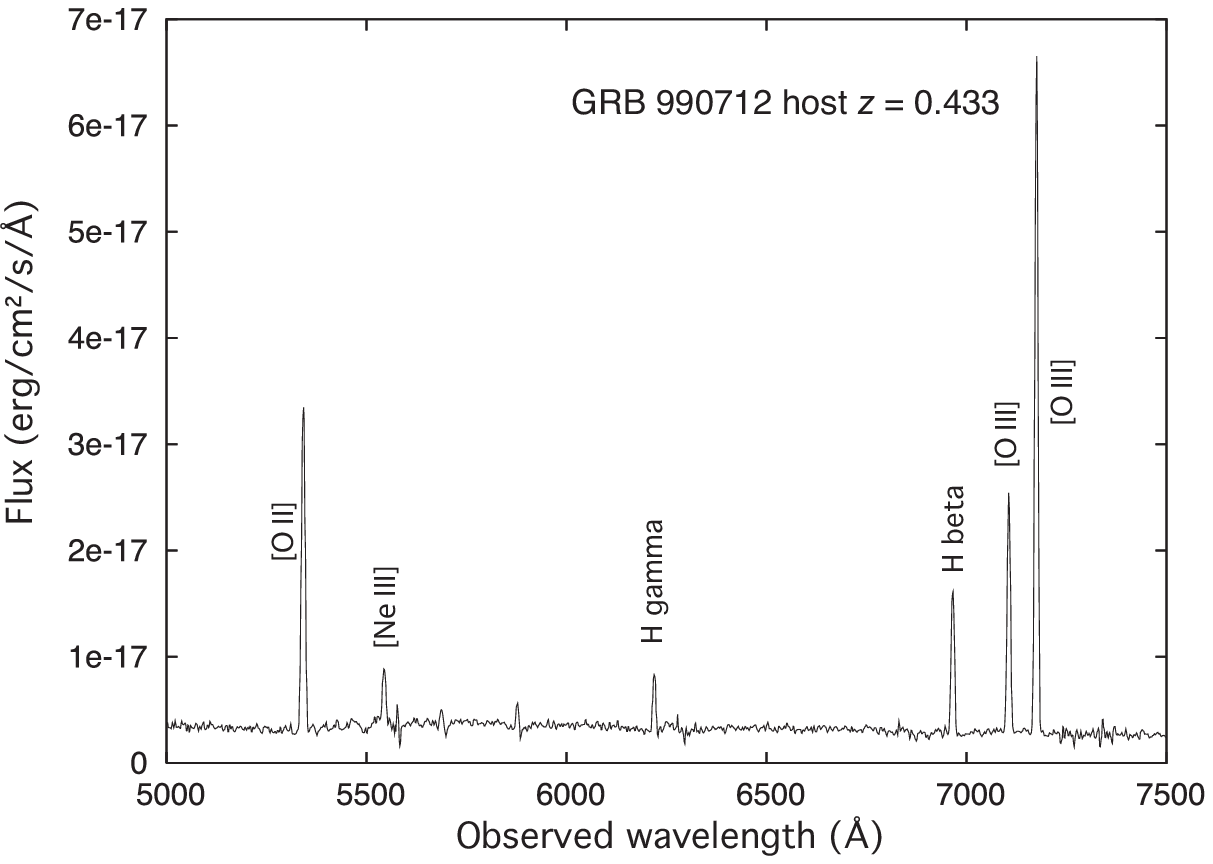}}
\begin{center}
 \caption{Two typical GRB host galaxies. {\it Left panel}: optical-NIR spectral energy distribution of the host of GRB~030329 (Gorosabel et al.\ 2005). The solid curve is the best-fit template with metallicity 1/5 solar. {\it Right panel}: optical integrated-light spectrum of the host of GRB~990712 (K{\"u}pc{\"u} Yolda{\c s} et al.\ 2006).}
   \label{fig5}
\end{center}
\end{figure}

\section{GRB host galaxies}

More than 50 galaxies associated with GRB events have been discovered so far. The largest public database GHostS (GRB Host Studies; www.grbhosts.info) lists many important observed parameters for most of them. These include multi-band photometry and optical emission-line fluxes. GHostS is also an interactive tool offering features such as SDSS and DSS sky viewing through Virtual Observatory services (Savaglio et al.\ 2007).

\begin{figure}
\centerline{\includegraphics[scale = .4]{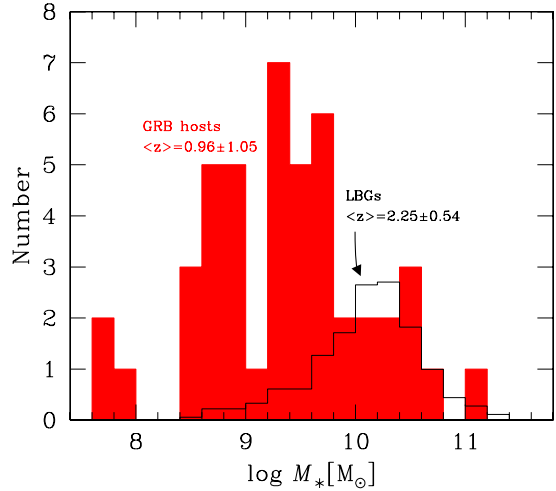}}
\begin{center}
 \caption{Stellar mass histogram of GRB hosts (filled histogram), mean redshift and dispersion $z=0.96\pm1.05$. For comparison, the empty histogram represents LBGs  (redshift interval $1.3<z<3$; Reddy et al. 2006) normalized to the GRB host histogram for $M_\ast>10^{10}$ M$_\odot$.}
   \label{fig6}
\end{center}
\end{figure}

The typical GRB host (Figure~\ref{fig5}) is an intermediate-redshift, sub-luminous, blue, star-forming galaxy (Le Floc'h et al.\ 2003; Christensen et al.\ 2004; Prochaska et al.\ 2004). The largest sample of GRB hosts studied so far contains \nhosts\ galaxies (Savaglio, Glazebrook, Le Borgne 2008). Its redshift distribution is shown in Figure~\ref{z_hist}. The average redshift and dispersion are $<z>= 1.0 \pm 1.0$. Among other parameters, Savaglio et al.\ (2008) measured the SFR and stellar mass of the sample. The histogram distribution of the latter is shown in Figure~\ref{fig6}. 

\begin{figure}
\centerline{\includegraphics[scale = .45]{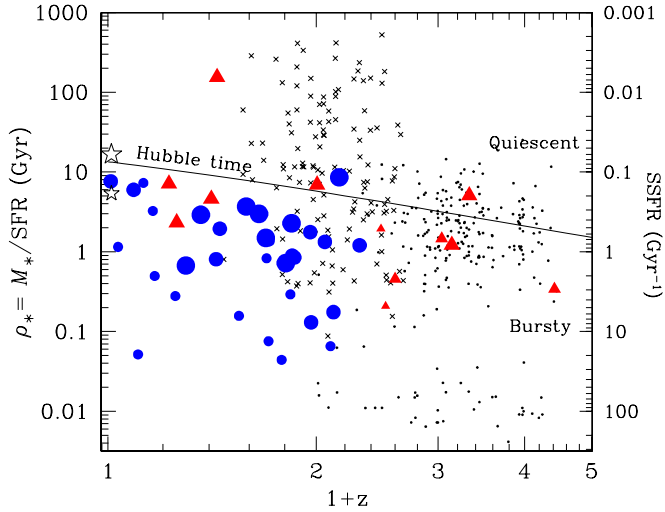}}
\begin{center}
 \caption{Growth time scale $\rho_\ast=M_\ast/SFR$ (left y-axis) or specific star formation rate SSFR (right y-axis) as a function of redshift. Filled circles and triangles are GRB hosts with SFRs measured from emission lines and UV luminosities, respectively.  Small, medium and large symbols are hosts with $M_\ast \leq 10^{9.0}$ M$_\odot$, $10^{9.0}$ M$_\odot < M_\ast \leq 10^{9.7}$ M$_\odot$, and $M_\ast> 10^{9.7}$ M$_\odot$, respectively. The curve is the Hubble time as a function of redshift, and indicates the transition from bursty to quiescent mode for galaxies. Crosses are field galaxies at $0.5<z<1.7$ (Juneau et al.\ 2005; Savaglio et al.\ 2005). Dots are LBGs at $1.3\lsim z\lsim 3$ (Reddy et al.\ 2006). The big and small stars at zero redshift represent the Milky Way and the Large Magellanic Cloud, respectively.}
   \label{fig7}
\end{center}
\end{figure}

The specific star formation rate ($SSFR=SFR/M_\ast$) or its inverse, the growth time scale ($\rho_\ast = M_\ast/SFR$) as a function of redshift is shown in Figure~\ref{fig7}. This shows how active a galaxy is. In particular $\rho_\ast$ gives the time that a galaxy needs to form the observed stellar mass, if the measured SFR has been constant over its entire life. In Figure~\ref{fig7} we also mark the age of the universe (Hubble time) as a function of redshift. Galaxies above the line can be considered quiescent. Most GRB hosts are below the line, half of them have SSFR above 0.8 Gyr$^{-1}$, or $\rho_\ast<800$ Myr. 

Metallicities in the HII regions of GRB hosts are measured with different methods, using emission-line fluxes. Results for 17 GRB hosts at $z<1.0$ are shown in Figure~\ref{fig4} (Savaglio et al.\ 2008). The 5 metallicities derived using the electron temperature $T_e$ and O3N2 methods, give a more reliable result. The other metallicities, mainly measured with the notoriously problematic $R_{23}$ calibrator, which gives two solutions (lower and upper branch; Kewley \& Ellison 2008), are more uncertain. Either solution, combined with all the other metallicities (including those in GRB-DLAs), does not reveal a redshift evolution of metallicities in GRB hosts in the interval $0<z<6.3$ (Figure~\ref{fig4}).

\section{Discussion}
   
The main, still unsolved, question about GRB hosts is whether they represent a fair sample of the whole star-forming galaxy population, or they are a distinct population of galaxies. GRB hosts are generally small, star-forming galaxies detected at any redshift, up to $z=6.3$. The mean stellar mass (measured mainly for the low-$z$ subsample) is a few times above $10^9$ M$_\odot$. That is the stellar mass of the Large Magellanic Cloud. The mean SFR is 5 times higher than the LMC (Savaglio et al.\ 2008). Metallicities of the hosts, measured from absorption lines in the afterglow spectra at $z>2$ or from emission lines in the host spectra at $z<1.0$, do not indicate a clear redshift evolution (Figure~\ref{fig4}), with values mainly between solar and 1/10 solar.  This suggests that the chemical enrichment is not a parameter characterizing the GRB host population.

It is clear that small star-forming galaxies are the most common galaxies that existed in the entire history of the universe. For instance, the star formation history of the universe, studied for different galaxy stellar-mass bins, indicates that small galaxies dominated for redshift $z<2$, where massive galaxies experienced a fast decline (Juneau et al.\ 2005). GRB hosts are low stellar-mass star-forming galaxies and can provide a very efficient way to understand galaxy formation and evolution in the most active phase of the universe. Future missions will give the possibility to study them in much larger quantities.

\begin{acknowledgement}
We thank the conference organizers for the outstanding organization.
The author thanks Karl Glazebrook and Damien Le Borgne for the very fruitful and long-lasting collaboration.

\end{acknowledgement}




\end{document}